\documentclass[pre,superscriptaddress,twocolumn,showpacs]{revtex4-1}
\usepackage{amsmath}
\usepackage{amssymb}
\usepackage{graphicx}

\begin{document}

\title{Disentangling geometric and dissipative origins
	   of negative Casimir entropies}

\author{Stefan Umrath}
\email{stefan.umrath@physik.uni-augsburg.de}
\affiliation{Institut f{\"u}r Physik, Universit{\"a}t Augsburg,
       Universit{\"a}tsstra{\ss}e 1, D-86135 Augsburg, Germany}
\author{Michael Hartmann}
\affiliation{Institut f{\"u}r Physik, Universit{\"a}t Augsburg,
       Universit{\"a}tsstra{\ss}e 1, D-86135 Augsburg, Germany}
\author{Gert-Ludwig Ingold}
\email{gert.ingold@physik.uni-augsburg.de}
\affiliation{Institut f{\"u}r Physik, Universit{\"a}t Augsburg,
       Universit{\"a}tsstra{\ss}e 1, D-86135 Augsburg, Germany}
\author{Paulo A. Maia Neto}
\affiliation{Instituto de F{\'i}sica, UFRJ, CP 68528, Rio de Janeiro RJ
       21941-909, Brazil}

\date{\today}

\begin{abstract}

Dissipative electromagnetic response and scattering geometry are potential
sources for the appearance of a negative Casimir entropy. We show that the
dissipative contribution familiar from the plane-plane geometry appears also in
the plane-sphere and the sphere-sphere geometries and adds to the negative
Casimir entropy known to exist in these geometries even for perfectly
reflecting objects. Taking the sphere-sphere geometry as an example, we carry
out a scattering-channel analysis which allows to distinguish between the
contributions of different polarizations. We demonstrate that dissipation and
geometry share a common feature making possible negative values of the Casimir
entropy. In both cases there exists a scattering channel whose contribution to
the Casimir free energy vanishes in the high-temperature limit. While the
mode-mixing channel is associated with the geometric origin, the transverse
electric channel is associated with the dissipative origin of the negative
Casimir entropy. By going beyond the Rayleigh limit, we find even for large
distances that negative Casimir entropies can occur also for Drude-type metals
provided the dissipation strength is sufficiently small.

\end{abstract}


\maketitle

\section{Introduction}

The specific heat of a system coupled infinitely weakly to a thermal heat bath
always has to be positive. This statement derives directly from the fact that the
specific heat is proportional to the variance of the system energy. By
appropriately integrating the specific heat, it follows that also the entropy
has to be positive. Nevertheless, there are a number of situations in which the
existence of a negative specific heat or a negative entropy is discussed.

Such a situation immediately arises when the coupling between the system and
the heat bath is no longer infinitely weak \cite{Haenggi2008,Ingold2009a}.
Then, the specific heat of the system can be defined as the difference between
the specific heat of system plus heat bath on the one hand and the specific heat
of the heat bath alone on the other hand. While there is no reason why the
difference of two positive numbers has to be positive, one would typically
expect that coupling additional degrees of freedom will increase the specific
heat. The difference of specific heats just introduced is therefore expected to
be positive. The same reasoning applies to the entropy.

A notable exception is the damped free particle, for which under appropriate
circumstances, the specific heat can become negative
\cite{Haenggi2008,Ingold2012,Spreng2013,Adamietz2014}. An interesting feature
of this system consists in the fact that its properties depend on the
dimensionless ratio $k_BT/\hbar\gamma$ where $T$ is the temperature and
$\gamma$ the damping constant while $k_B$ and $\hbar$ are the Boltzmann
constant and the Planck constant, respectively. As a consequence and in
contrast to common expectations, decreasing the damping constant $\gamma$ will
render the free particle more classical and shift the transition to the
quantum behavior to lower temperatures. A negative specific heat in the sense
just discussed is also found in certain systems in condensed-matter physics and
related fields
\cite{Florens2004,Zitko2009,Campisi2009,Campisi2010,Sulaiman2010}.

Another important case where negative entropies give rise to considerable
interest is the Casimir effect which we will consider in more detail in the
present paper. Even though experimentally the Casimir force is much more
relevant than the Casimir entropy, the former contains a significant
zero-temperature contribution while the latter allows to focus on the effect of
thermal photons.

Suppose that we consider the Casimir interaction between two
objects. With each of the objects as well as with the combination of the two
objects we can associate an entropy as the difference between the entropy of
the electromagnetic field in the presence of the object(s) and the entropy of
the free electromagnetic field. The Casimir entropy is an interaction entropy
which then is obtained by subtracting the entropies of the two individual
objects from the entropy of the two objects. While individual entropies are
not finite, the Casimir entropy yields a finite value.

One potential source of a negative Casimir entropy is the coupling of the
electromagnetic field to the electrons in the scattering objects. These
electrons undergo a dissipative motion and therefore the electromagnetic
response of the objects is characterized by a damping constant $\gamma$. The
presence of dissipation inside the scatterers leads to a suppressed
reflectivity of transverse electric (TE) modes at low frequencies while the
scattering of transverse magnetic (TM) modes is not affected significantly.
As a consequence, in the plane-plane geometry a reduction of the Casimir force
\cite{Bostrom2000} and Casimir entropy by a factor of two is found at high
temperatures. The possibility of a negative Casimir entropy has been the
subject of an extensive debate over more than a decade \cite{Bezerra2002,
Bostrom2004,Bezerra2004,Brevik2004,Brevik2005,Svetovoy2005,Bezerra2008,
Ellingsen2009,Ingold2009b,Pitaevskii2010,Canaguier2010,Canaguier2010a,
Bordag2010,Weber2010,Zandi2010,Rodriguez2011,Ingold2015,Milton2015}.

A thorough understanding of the role of dissipation in the Casimir effect is of
considerable interest for the interpretation of experiments. Presently, there
still exists disagreement about whether the dissipative low-frequency part of
the electromagnetic response is relevant or not. While some experiments are
interpreted in terms of the dissipation-less plasma model
\cite{Decca2007,Banishev2013}, other experiments are interpreted in terms
of the Drude model which includes dissipation \cite{Sushkov2011,Garcia2012}.
At this point, it is useful to recall that a nonvanishing value of $\gamma$ is
required for the Lifshitz formula to be applicable \cite{Guerout2014}.

Most Casimir force measurements are carried out in the plane-sphere geometry
with a surface-to-surface distance between 0.1 and a few micrometers.
Depending on the experiment, the radius of the sphere or the radius of
curvature of a spherical segment ranges from $20\,\mu$m
\cite{Jourdan2009,Torricelli2011} to about 20\,cm \cite{Masuda2009}. The latter
case is rather close to a configuration of two parallel plates, where TE and TM
modes do not mix. However, with decreasing sphere radius, polarization mixing
becomes increasingly relevant and it is important to understand how the
occurrence of negative Casimir entropies depends on dissipation in the
plane-sphere and sphere-sphere geometry. 

The extension of the discussion of the negative Casimir entropy due to
dissipation beyond the plane-plane geometry is not the only purpose of this
paper. In fact, it is known since some time, that negative Casimir entropies
can arise for the plane-sphere \cite{Canaguier2010,Canaguier2010a,Zandi2010}
and the sphere-sphere \cite{Rodriguez2011} geometry even for perfect
conductors. An important question therefore concerns the relative status of the
negative Casimir entropies due to dissipation on the one hand and due to
geometry on the other hand.

Recently, it was shown that the negative Casimir entropy in the perfect metal
plane-sphere and sphere-sphere geometries can be traced back to the mixing
between the TE and TM modes of the electromagnetic field
\cite{Ingold2015,Milton2015}. This identification became possible by realizing
that negative Casimir entropies are most pronounced at large distances between
the scattering objects. Then, within a scattering formalism, it is sufficient
to consider one round-trip of electromagnetic waves between the scatterers and a
decomposition into the various scattering channels becomes possible.

In Refs.~\cite{Ingold2015,Milton2015} it was argued that in the large-distance
limit two Drude-type metal objects will not allow for a negative Casimir
entropy because the reflection of the TE modes is strongly suppressed and only
a positive contribution of TM modes is left. This regime can be attained for
any given damping strength by making the distance between the scattering
objects sufficiently large. On the other hand, for any given distance of the
objects, one can make the damping strength so small that a negative Casimir
entropy ensues.

In order to disentangle the contributions to a negative Casimir entropy arising
from dissipation and geometry, we will go beyond the electric dipole
approximation, i.e. the Rayleigh limit \cite{Bohren2007} employed in
Ref.~\cite{Ingold2015}. While we will assume the distance $d$ between the
scattering objects to be large compared to the radii of the spherical objects
involved, we will allow for damping strengths so small that a dissipative
contribution to the negative Casimir entropy can occur in addition to the
geometric contribution.

In Sec.~\ref{sec:dissgeo}, we start by sketching the formalism required for the
polarization channel analysis introduced in Ref.~\cite{Ingold2015}. By
considering the Casimir entropy for the plane-plane, sphere-plane, and
sphere-sphere geometries, we then illustrate how negative Casimir entropies of
dissipative and geometric origin manifest themselves in the temperature
dependence of the Casimir entropy.  In Sec.~\ref{sec:MieCoefficients}, the
contributions from the various scattering channels are analyzed within the
dipole approximation but beyond the Rayleigh limit. The approximation of a
single scattering round-trip between the two scattering objects allows us to
disentangle dissipation and geometry as sources of a negative Casimir entropy
and at the same time to identify common features. In
Sec.~\ref{sec:beyond_dipole}, we go beyond the single round-trip approximation
and demonstrate that repeated scattering round-trips and, for perfect
conductors, higher multipoles tend to suppress negative Casimir entropies. In
Sec.~\ref{sec:conclusions} we present our conclusions. An appendix collects
some formulae needed for the evaluation of the Casimir entropy within the
scattering formalism.

\section{Dissipative and geometrical influence on Casimir thermodynamics}
\label{sec:dissgeo}

The Casimir entropy is obtained from the Casimir free energy $\mathcal{F}$
by means of the usual thermodynamic relation
\begin{equation}
\label{eq:entropy}
\mathcal{S}=-\frac{\partial \mathcal{F}}{\partial T}\,.
\end{equation}
Within the multiple-scattering theory \cite{Jaekel1991,Genet2003,Lambrecht2006,
Emig2007}, the Casimir free energy can be expressed as
\begin{equation}
\label{eq:free_energy}
\mathcal{F} = \frac{k_B T}{2}\sum_{n=-\infty}^\infty\mathrm{Tr}\left(
	      \ln\big[1-\mathcal{M}(\vert\xi_n\vert)\big]\right)
\end{equation}
with the Matsubara frequencies $\xi_n =2\pi nk_B T/\hbar$. The matrix
$\mathcal{M}$ describes round-trips of electromagnetic waves between all
involved scatterers at imaginary frequencies $\mathrm{i}\xi_n$. 

As we will focus on the sphere-sphere and plane-sphere geometries, the trace in
(\ref{eq:free_energy}) can be taken in a spherical wave basis with the angular
part given by spherical harmonics of degree $\ell$ and order $m$. In view
of the axial symmetry of the geometries considered here, $m$ is conserved during
the scattering process. The round-trip operator in a corresponding subspace
\begin{equation}
\mathcal{M}^{(m)} = \mathcal{R}_1^{(m)} \mathcal{T}_{12}^{(m)} 
                    \mathcal{R}_2^{(m)} \mathcal{T}_{21}^{(m)}
\label{eq:roundtrip}
\end{equation}
can be expressed as product of translation operators $\mathcal{T}_{ij}$ from the
reference frame of object $j$ to that of object $i$ and reflection operators
$\mathcal{R}_i$ associated with object $i$. For reference, explicit expressions
for the sphere-sphere geometry involving Mie coefficients and spherical wave
translation formulas are given in the appendix.

For large distances between the scattering objects, all matrix elements of the
round-trip operator $\mathcal{M}$ will be much smaller than one. The Casimir free energy
(\ref{eq:free_energy}) can then be linearized,
\begin{equation}
\label{eq:single_scattering}
\mathcal{F} \approx -\frac{k_B T}{2}\sum_{n=-\infty}^\infty\sum_{m=-\infty}^\infty
                    \mathrm{Tr}\mathcal{M}^{(\vert m\vert)}(\vert\xi_n\vert)\,,
\end{equation}
thereby retaining only contributions from single round-trips. Within this
approximation it is possible to decompose the Casimir free energy $\mathcal{F}$
and the Casimir entropy $\mathcal{S}$ into contributions from the different
scattering channels and thereby to gain physical insight
\cite{Ingold2015,Milton2015}. Of particular interest will be the channel where
the TE polarization is conserved during the round-trip and the
polarization-mixing channel where each of the two objects scatters a different polarization. 

Furthermore, for sufficiently large separations $d/R\gtrsim 20$, the dipole
approximation applies to the Casimir entropy and we can restrict ourselves to
$\ell=1$ and $\vert m\vert=0,1$ (see Figs.~\ref{fig:SRT_exact_PEC} and
\ref{fig:SRT_exact_Drude} below). We will make use of the single round-trip
dipole approximation in the present section as well as in the following
Sec.~\ref{sec:MieCoefficients}. Its range of validity will be discussed in
Sec.~\ref{sec:beyond_dipole}.

In order to obtain a first idea of the interplay between the dissipative and
geometric contributions to the negative Casimir entropy in the plane-sphere
and sphere-sphere geometries, we compare with the plane-plane geometry where
the negative Casimir entropy is solely of dissipative origin. Apart from the
special case of perfectly conducting metals (P), we will consider non-magnetic
Drude-type (D) metals with the frequency-dependent permittivity
\begin{equation}
 \label{eq:epsDrude}
 \varepsilon(\omega) = 1-\frac{\omega_P^2}{\omega(\omega+i\gamma)}\,.
\end{equation}
Here, $\gamma$ and $\omega_P$ are the damping constant and the plasma
frequency, respectively, corresponding to a dc conductivity $\sigma_0
=\omega_P^2/\gamma$. While $\omega_P$ determines the frequency scale beyond
which the metal becomes transparent, a finite value of $\sigma_0$ results
in a vanishing reflectivity of the TE modes at zero frequency. 

\begin{figure}
\centering
\includegraphics[width=\columnwidth]{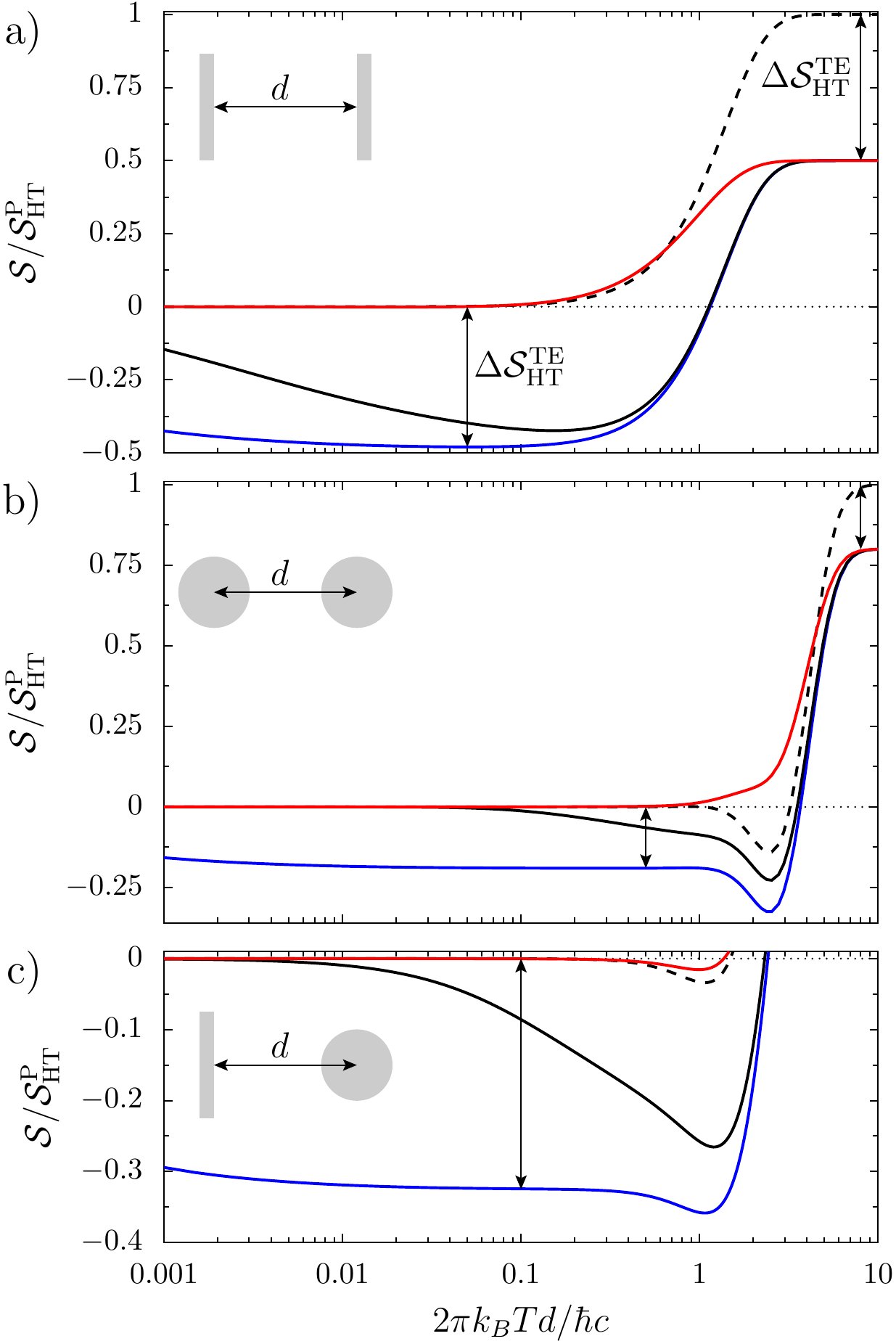}
\caption{The Casimir entropy $\mathcal{S}$ scaled by its high-temperature
  value $\mathcal{S}_\text{HT}^\text{P}$ for perfect conductors is shown as
  a function of temperature for the (a) plane-plane geometry, (b) sphere-sphere
  geometry, and (c) plane-sphere geometry. While the dashed lines represent the
  entropy for perfectly conducting objects, the solid lines correspond to
  Drude-type conductors with the damping constant $\gamma d/c=10^{-2}, 10^2,$
  and $10^{4}$ increasing from bottom to top and the plasma frequency given by
  $\omega_P d/2\pi c=400$. The sphere radii are chosen as $R=d/20$. The
  arrows indicate the effect of the missing zero-frequency contribution to the
  Casimir entropy for Drude-type metals with respect to perfect conductors.}
\label{fig:PP_SP_SS_entropies}
\end{figure}

Figure~\ref{fig:PP_SP_SS_entropies} shows the temperature dependence of the
Casimir entropy for the three geometries indicated by the insets. The entropy
is scaled by its high-temperature value $\mathcal{S}_\text{HT}^\text{P}$ for
perfectly conducting metals. A dimensionless temperature is defined by means of
the distance $d$ between the reference frames of the two scattering objects.
The three solid lines correspond to Drude-type metals with $\gamma d/c=
10^{-2}, 10^2,$ and $10^4$ increasing from bottom to top and $\omega_P d/2\pi
c=400$. Here, $c$ is the speed of light. The lower (blue) curve and the upper
(red) curve thus correspond to good and bad conductors, respectively. As a
reference, the case of a perfectly conducting metal is depicted as dashed line.

We start by recalling in Fig.~\ref{fig:PP_SP_SS_entropies}a the main features
of the Casimir entropy for Drude-type metals in the plane-plane geometry.  It
is well known that a finite dc conductivity suppresses the zero-frequency
contribution of the TE modes, resulting in a reduction
$\Delta\mathcal{S}_\text{HT}^\text{TE}$ of the Casimir entropy at high
temperatures by a factor of two with respect to the case of an infinite dc
conductivity \cite{Bostrom2000}. The arrows in
Fig.~\ref{fig:PP_SP_SS_entropies} visualize the effect of this missing
zero-frequency contribution to the Casimir entropy.

Since for perfectly conducting metals the Casimir entropy decreases with
decreasing temperature, the missing zero-frequency contribution of the TE modes
exceeds their contribution at finite temperatures. As a consequence, at
sufficiently low temperatures, the Casimir entropy for Drude-type metals
becomes negative, provided the dc conductivity is sufficiently large.  For the
smallest value of the damping constant $\gamma$ in
Fig.~\ref{fig:PP_SP_SS_entropies}a the entropy curve over a large temperature
range ressembles closely that for a perfect conductor except for the entropy
shift $\Delta\mathcal{S}_\text{HT}^\text{TE}$.

The Casimir entropy approaches its vanishing zero-temperature
value only when the temperature falls below a temperature of the order of
$\hbar\gamma/k_B$ \cite{Ingold2009b}. The middle (black) curve therefore
comes close to zero at much higher temperatures than the lower (blue) curve.
This behavior is reminiscent of the damped free particle discussed in the 
introduction where the temperature scale is also set by the damping constant
$\gamma$.

A reduction of the Casimir entropy at high temperatures as well as negative
values at lower temperatures with a crossover to vanishing Casimir entropy on a
temperature scale given by $\gamma$ are also visible for the sphere-sphere
geometry in Fig.~\ref{fig:PP_SP_SS_entropies}b and the plane-sphere geometry in
Fig.~\ref{fig:PP_SP_SS_entropies}c. Since the two geometries share the same
qualitative features, we restrict the latter figure to negative values of the
Casimir entropy. In both figures, the sphere radii are chosen as $R=d/20$ to
ensure the validity of the single round-trip dipole approximation.

There are quantitative differences in the temperature dependence of the Casimir
entropy for the three geometries. While the temperature scale on which the
Casimir entropy approaches zero at low temperatures clearly increases with
increasing damping constant $\gamma$, its value depends on the scattering
geometry.  In addition, the reduction of the Casimir entropy at high
temperatures amounts only to 1/3 for the plane-sphere geometry and 1/5 for the
sphere-sphere geometry instead of 1/2 for the plane-plane geometry. This
reduction is a consequence of the fact that for a perfectly conducting sphere
the reflectivity of the TE modes is only one half of the reflectivity of the TM
modes.

Besides the features already present in the plane-plane geometry, an additional
dip in the Casimir entropy appears for the sphere-sphere and plane-sphere
geometries at temperatures where $2\pi k_BTd/\hbar c$ is of order one. This
feature is of geometric origin and survives even in the absence of dissipation
as can be seen from the dashed lines in Figs.~\ref{fig:PP_SP_SS_entropies}b and
c. Its position depends on the geometry and it is more pronounced in the
sphere-sphere geometry.

While for a damping constant as large as $\gamma=10^4c/d$, i.e. for the upper
(red) solid curves, the Casimir entropy still becomes negative in the
plane-sphere geometry, it remains positive for all temperatures in the
sphere-sphere geometry.  Nevertheless, even there a remnant of the dip is
visible. However, for sufficiently large damping constant $\gamma$, the
contribution to the Casimir entropy involving TE modes will disappear
\cite{Ingold2015,Milton2015}. For such strong damping, the electrons will
suffer many collisions during an oscillation period of the electromagnetic
field for all relevant frequencies. The total irradiating power in the TE modes
will then be dissipated, thereby suppressing the reflection of these modes.

\section{Channel analysis beyond the Rayleigh limit}
\label{sec:MieCoefficients}

The temperature dependence of the Casimir entropy displayed in
Figs.~\ref{fig:PP_SP_SS_entropies}b and c suggests that the negative Casimir
entropy can be separated into contributions of dissipative and of geometric
origin. This is particularly clear for the sphere-sphere geometry for which we
will demonstrate in the second half of this section that the expected
separation is indeed possible in terms of different scattering channels. Even
though we will be able to disentangle geometric and dissipative parts,
Figs.~\ref{fig:PP_SP_SS_entropies}b and c also show that the geometric
contribution depends on the damping constant appearing in the Drude-type
permittivity (\ref{eq:epsDrude}). In order to analyze this dependence, we will
need to take a closer look at the scattering properties of Drude-type metal
spheres.

For the sphere-sphere geometry, the translation operators $\mathcal{T}$
appearing in (\ref{eq:roundtrip}) restrict the relevant wave numbers to small
values with $kd\lesssim 1$. This cut-off is caused by the modified spherical
Bessel functions present in (\ref{eq:TPPp}). For large distances $d\gg R$, one
may therefore apply the Rayleigh limit, where only dipole scattering is
relevant.  For perfect metal spheres, it will be sufficient to consider only
the leading term in the expansion of the $\ell=1$ Mie coefficients for TM modes
\begin{equation}
\label{eq:mie_pTM}
a^{\rm P}_1 = -\frac{2}{3}(kR)^3+\frac{1}{5}(kR)^5+\mathcal{O}\left(k^6\right)\,,
\end{equation}
as well as for TE modes
\begin{equation}
\label{eq:mie_pTE}
b^{\rm P}_1 =\frac{1}{3}(kR)^3+\frac{1}{5}(kR)^5+\mathcal{O}\left(k^6\right)\,.
\end{equation}
Here, TM and TE scattering amplitudes are of the same order with a
relative factor of two alluded to in the previous section.

In contrast, for Drude-type metal spheres, the Mie coefficients for TM modes
with $\ell=1$
\begin{equation}
\label{eq:mie_drudeTM}
a^{\rm D}_1 = -\frac{2}{3}(kR)^3 + \frac{2c}{\sigma_0R}(kR)^4+\mathcal{O}(k^5)
\end{equation}
and for TE modes
\begin{equation}
\label{eq:mie_drudeTE}
\begin{aligned}
b^{\rm D}_1 &= \frac{R \sigma_0}{45 c}(kR)^4\\
 &\quad - \frac{1}{45}\left[\frac{2}{21}\left(\frac{\sigma_0R}{c}\right)^2
              +\frac{\sigma_0}{\gamma}\right](kR)^5+ \mathcal{O}(k^6)\,,
\end{aligned}
\end{equation}
differ in the power of $kR$ in the leading term. The scattering of TE modes
thus becomes negligible in the Rayleigh limit. However, the prefactor of the
leading term in (\ref{eq:mie_drudeTE}) indicates the appearance of a new
scale. As the comparison of (\ref{eq:mie_drudeTM}) and (\ref{eq:mie_drudeTE})
shows, the scattering of TE modes can only be neglected if
\begin{equation}
\label{eq:TE_condition}
\frac{R \sigma_0}{30 c} \ll \frac{d}{R}\,.
\end{equation}
The applicability of the Rayleigh limit thus also depends on the dc conductivity
$\sigma_0$. For a given dc conductivity, it is always possible to reach the 
large-distance limit where the scattering of TE modes can be neglected. On the
other hand, for a given distance, the Rayleigh limit ceases to hold if the 
dc conductivity becomes too large. 

A physical interpretation of the condition (\ref{eq:TE_condition}) can be given
in terms of the quasi-static skin depth
\begin{equation}
\delta(\omega) = \left(\frac{2c^2}{\sigma_0 \omega}\right)^{1/2}.
\end{equation}
Scattering of TE modes can thus only be neglected provided the sphere is
sufficiently small compared to the skin depth,
\begin{equation}
R \ll \sqrt{15}\delta\,,
\end{equation}
in the relevant frequency range. Then, no eddy currents can be induced and
the sphere becomes transparent to the TE modes.

\begin{figure}
\centering
\includegraphics[width=\columnwidth]{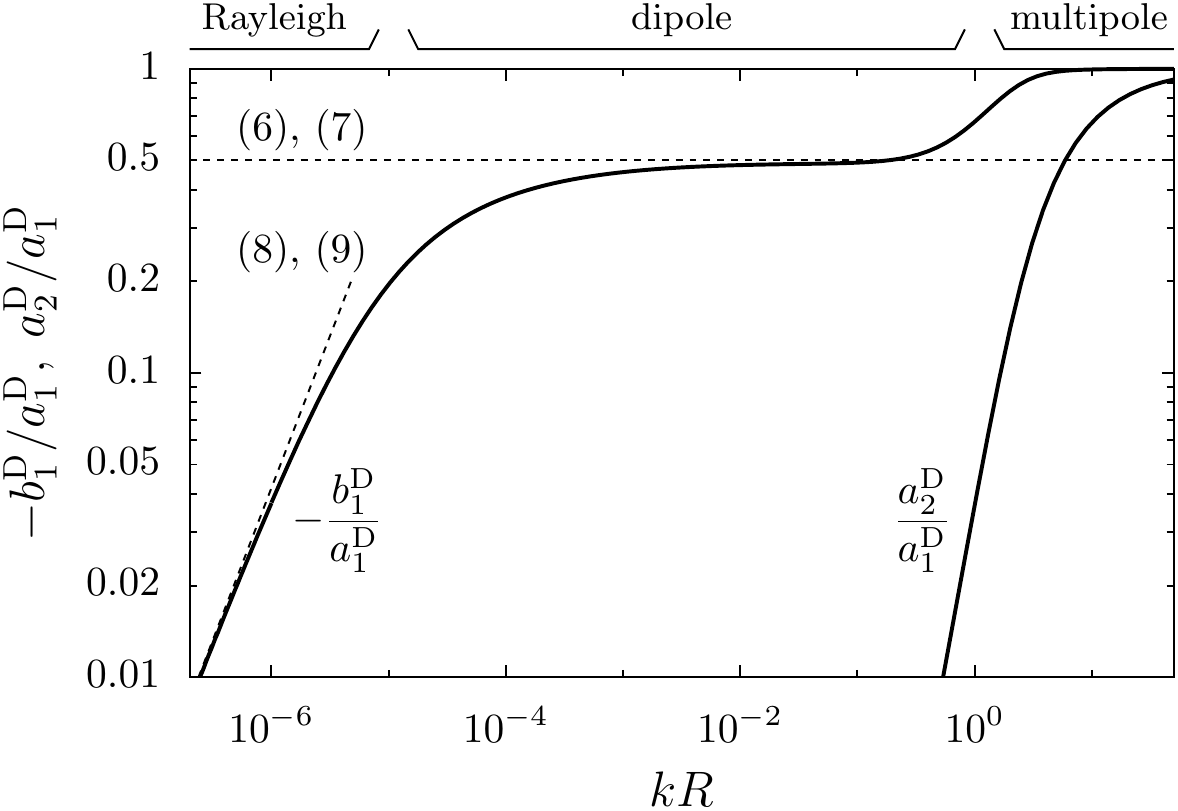}
\caption{The ratios of Mie coefficients $-b_1^\text{D}/a_1^\text{D}$ and
  $a_2^\text{D}/a_1^\text{D}$ as a function of the wave number $k$ are shown
  for a Drude-type metal sphere with radius $R$. The plasma frequency and the
  damping constant are given by $\omega_P R/2\pi c=20$ and
  $\gamma/\omega_P=10^{-4}$. Three different regimes can be distinguished.
  In the Rayleigh regime, TE scattering is negligible, while in the dipole
  regime, TE and TM scattering are of comparable strength. In the multipole
  regime, the scattering of higher multipole waves becomes relevant. The
  dashed lines indicate the approximations valid in the first two regimes with
  references to the corresponding equations.}
\label{fig:MieCoefficients}
\end{figure}

The breakdown of the validity of the Rayleigh regime for a well conducting
metal sphere with $\sigma_0R/c=4\pi\times10 ^4$ is illustrated in
Fig.~\ref{fig:MieCoefficients}. The ratio of Mie coefficients
$-b_1^\text{D}/a_1^\text{D}$ for dipole scattering approaches at small wave
numbers the behavior predicted by the leading terms in (\ref{eq:mie_drudeTM})
and (\ref{eq:mie_drudeTE}). At larger wave numbers, the dipole regime is
entered where the ratio of the Mie coefficients $a_1^\text{D}$ and
$b_1^\text{D}$ approaches the value for perfectly conducting spheres given by
(\ref{eq:mie_pTM}) and (\ref{eq:mie_pTE}). Increasing the wave numbers even
further, the multipole regime is reached where first the scattering of TM modes
with $\ell=2$ described by the Mie coefficient $a_2^\text{D}$ becomes relevant.

An increase in the dc conductivity will reduce the Rayleigh regime and render
the dipole regime valid at even smaller frequencies. To ensure the validity of the Rayleigh limit, one thus has
to choose either bad conductors or go to large distances satisfying
(\ref{eq:TE_condition}). In the large-distance limit considered in
Refs.~\cite{Ingold2015} and \cite{Milton2015}, it was assumed that
(\ref{eq:TE_condition}) holds and that the TE modes are not scattered by
Drude-type metal spheres. Under this condition, it was possible to derive
analytical results for the Casimir entropy. Here, we will allow for an
arbitrary dc conductivity and thus have to resort to a numerical evaluation
of the Casimir entropy employing the full Mie coefficients given in the appendix.

In the large-distance limit $d\gg R$, where the linearized single round-trip
expression (\ref{eq:single_scattering}) for the Casimir free energy applies, we
can decompose the Casimir entropy into contributions arising from the
scattering channels characterized by the polarization on the two spheres. For
two channels the polarization on the spheres is conserved along the
round-trip: TM$\leftrightarrows$TM and TE$\leftrightarrows$TE. In the
following, we will refer to them as TM and TE channels, respectively. In
addition, two channels correspond to polarization mixing:
TM$\leftrightarrows$TE and TE$\leftrightarrows$TM. Because of the symmetry of
our geometry with identical spheres, the two polarization-mixing channels
provide equal contributions and we will only consider their sum.

\begin{figure}
\centering
\includegraphics[width=\columnwidth]{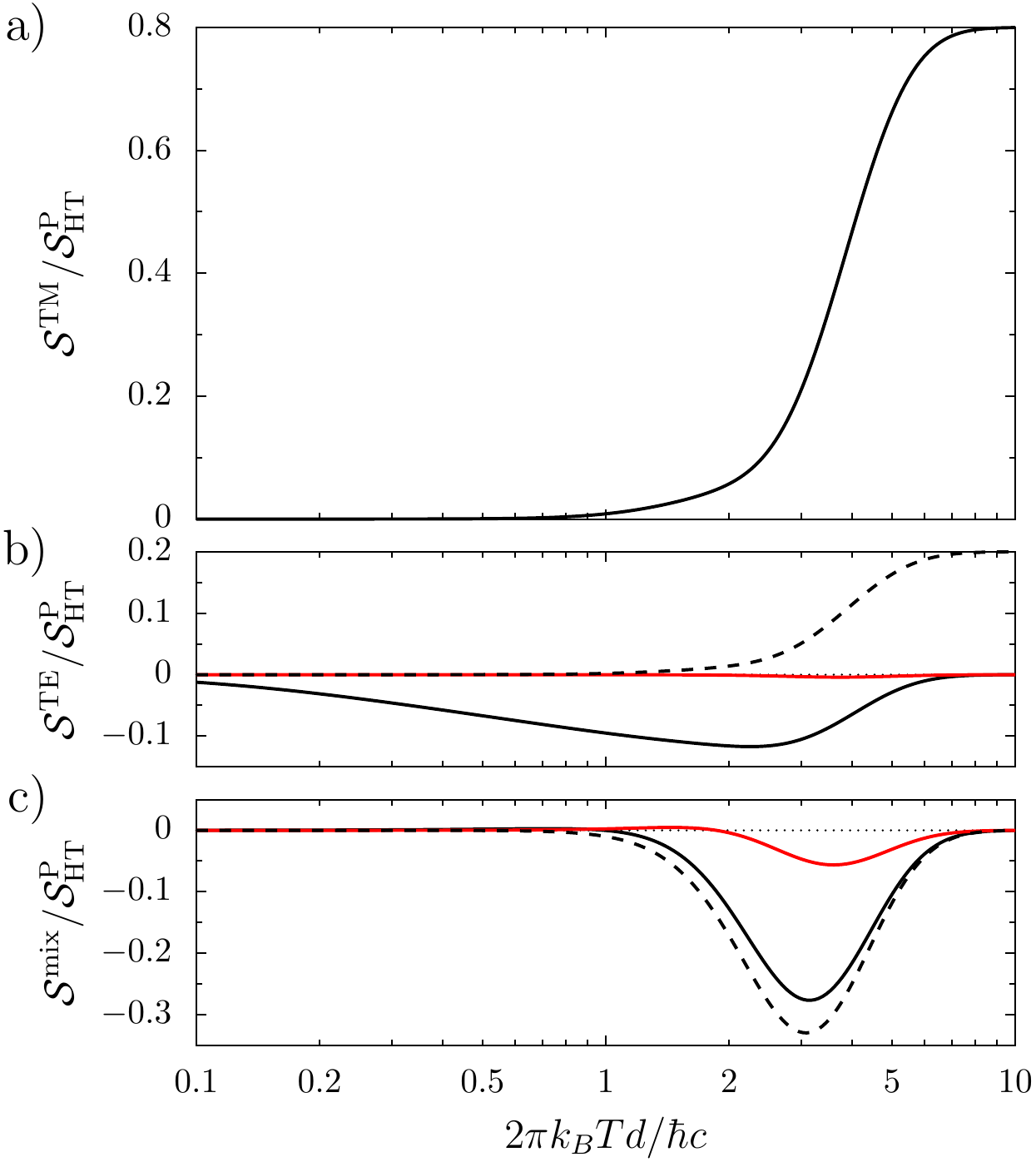}
\caption{The contributions from (a) the TM channel, (b) the TE channel, and
  (c) the polarization-mixing channels to the Casimir entropy are shown as
  function of temperature for a sphere-sphere geometry with $d/R=20$. All
  contributions are scaled by the total high-temperature Casimir entropy
  $\mathcal{S}^\text{P}_\text{HT}$ for perfectly conducting spheres. The
  parameters are chosen as in Fig.~\ref{fig:PP_SP_SS_entropies}, i.e. the dashed 
  lines correspond to perfect metal spheres while the solid black and red lines
  correspond to Drude-type metal spheres with $\gamma d/c=10^2$ and $10^4$,
  respectively, and $\omega_P d/2\pi c=400$. For the sake of clarity, panel (a)
  displays only one line because all other lines deviate by not more than one
  line width. To facilitate the comparison between the contributions of the
  three types of scattering channels, the scales on the vertical axes are
  chosen to be equal.}
\label{fig:spheresphere_entropy}
\end{figure}

Fig.~\ref{fig:spheresphere_entropy} displays the temperature dependence of the
three different contributions to the Casimir entropy scaled by the
high-temperature Casimir entropy $\mathcal{S}_\text{HT}^\text{P}$ for perfectly
conducting spheres. The parameters equal those used in
Fig.~\ref{fig:PP_SP_SS_entropies} but no curves for the highest dc conductivity
are shown.

In Fig.~\ref{fig:spheresphere_entropy}a, the contribution of the TM channel is
depicted. We only show the curve corresponding to $\gamma d/c=10^2$ but omit
the curves for $\gamma d/c=10^4$ and for perfect conductors. According to
(\ref{eq:mie_pTM}) and (\ref{eq:mie_drudeTM}), the reflection coefficient for
TM modes depends only very weakly on the dc conductivity and, in fact, the
other curves would lie within not more than one line width. As discussed
before, at high temperatures the TM channel contributes 4/5 of the total
Casimir entropy. The vertical axes in Fig.~\ref{fig:spheresphere_entropy} are
drawn to scale so that a comparison between the three contributions is
facilitated.  While the TM channel largely dominates at high temperatures, this
is no longer the case at low temperatures, thus opening the possibility for
negative Casimir entropies.

The temperature dependence of the TE channel shown in
Fig.~\ref{fig:spheresphere_entropy}b displays the same features as the Casimir
entropy in the plane-plane geometry depicted in
Fig.~\ref{fig:PP_SP_SS_entropies}a once the contribution arising from the TM
modes is subtracted. We recognize the shift between the Casimir entropies for
perfectly conducting spheres and Drude-type metal spheres at high temperatures
due to the vanishing reflectivity for the TE modes at zero frequency. For not
too small dc conductivity, i.e. the solid black curve, the contribution to the
Casimir entropy becomes negative and goes to zero on a temperature scale
proportional to $\gamma$. For bad conductors, i.e. the red solid curve, the
contribution of the TE channel almost vanishes, in agreement with our
discussion of the Rayleigh limit. The scattering channel analysis thus confirms
that the TE modes are not only responsible for the dissipative contribution to
the negative Casimir entropy in the plane-plane geometry, but also in the
sphere-sphere geometry. It is worth mentioning that the same conclusion can be
shown to hold true for the plane-sphere geometry.

Polarization-mixing channels do not exist in the plane-plane geometry but
contribute to the Casimir entropy in the plane-sphere and sphere-sphere
geometries. The temperature dependence for the latter case is shown in
Fig.~\ref{fig:spheresphere_entropy}c. From Refs.~\cite{Ingold2015} and
\cite{Milton2015} it is already known that the polarization-mixing
channels are associated with the geometric origin of a negative
Casimir entropy. Here, we can now study the dependence of this feature as a
function of the Drude damping constant $\gamma$.

As the curves in Fig.~\ref{fig:spheresphere_entropy}c show, the negative
Casimir entropy tends to be increasingly suppressed with increasing damping
constant $\gamma$.  In constrast to the TE channel in
Fig.~\ref{fig:spheresphere_entropy}b, a continuous transition exists from
perfectly conducting spheres, where the negative Casimir entropy is most
pronounced, to bad Drude-type metal spheres, where the mode mixing is
significantly suppressed. Reducing the dc conductivity even further would
ultimately result in a vanishing geometric contribution to a negative Casimir
entropy \cite{Ingold2015,Milton2015}.

\begin{figure}
\includegraphics[width=\columnwidth]{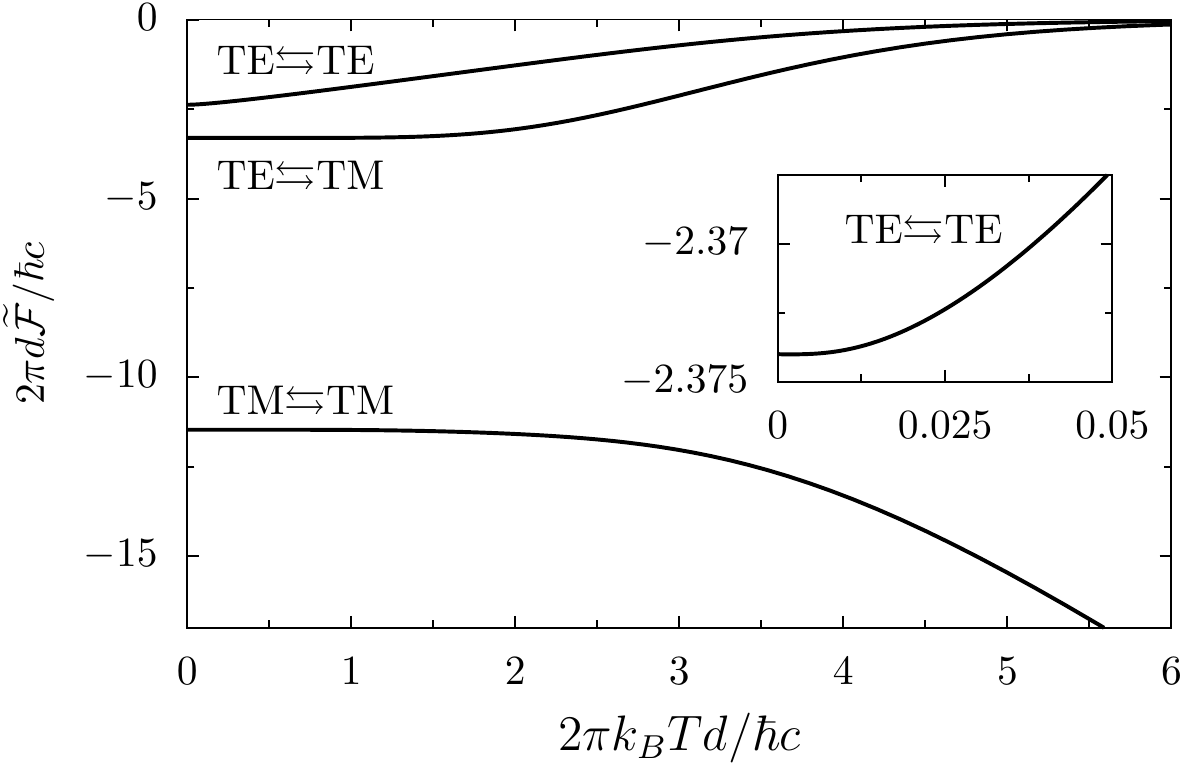}
\caption{The contributions of the different scattering channels to the
  scaled free energy $\widetilde{\mathcal{F}}=(d/R)^6\mathcal{F}$ are shown
  as a function of temperature. The distance between the two Drude-type
  metal spheres is given by $d/R=20$ and the material parameters are
  $\gamma d/c=10$ and $\omega_P d/2\pi c=400$. The inset displays the
  low-temperature behavior of the TE channel and demonstrates that the 
  corresponding contribution to the Casimir entropy vanishes in the
  zero-temperature limit.}
\label{fig:spheresphere_freeE}
\end{figure}

We conclude that the geometric and dissipative contributions to a negative
Casimir entropy can clearly be distinguished by means of a scattering-channel
analysis provided the single round-trip condition is met, i.e. $d\gg R$.
Nevertheless, the two types of channels share a common feature as can be seen
from Fig.~\ref{fig:spheresphere_freeE} where the contributions of the different
scattering channels to the Casimir free energy is shown as a function of
temperature.

The contributions of both, TE and polarization-mixing channels vanish in the
high-temperature limit. This behavior is in clear contrast to the usual
behavior found for the TM channel which yields a negative contribution
decreasing linearly with temperature at high temperatures. In view of
(\ref{eq:entropy}) and the fact that the contributions to the Casimir free
energy at zero temperature are negative, the temperature dependence of the
first two channels necessarily implies a negative contribution to the Casimir
entropy in some temperature range. 

This common feature should not obscure their different physical origins. For
the TE channel, the contribution to the Casimir entropy at high temperatures
vanishes because of the vanishing zero-frequency reflectivity. In contrast, for
the polarization-mixing channel, no contribution to the Casimir entropy can
occur at zero frequency where the fields corresponding to the two modes are
either purely electric or purely magnetic in any of the two coordinate systems
centered at each sphere.  The difference in the physical mechanism manifests
itself in Figs.~\ref{fig:spheresphere_entropy}b and c, where we observe that
the transition from a Drude-type metal to a perfect conductor occurs
continuously for the polarization-mixing channel while it does not for the TE
channel.

\section{Sphere-sphere Casimir entropy beyond the dipole approximation}
\label{sec:beyond_dipole}

So far, our discussion of the Casimir entropy has relied on the large-distance
assumption $d\gg R$. Only then it suffices to consider single round-trips and
the contributions of dissipative and geometric origin can be added. When the
distance between the two spheres is decreased, it becomes necessary to go
beyond the dipole approximation and to allow for scattering processes 
involving more than a single round-trip.

In this section, we first discuss the role of multiple round-trip contributions
for perfectly conducting spheres and thus concentrate on the geometrically
induced negative Casimir entropy. In a second step, we add dissipation by
considering Drude-type metal spheres.

\begin{figure}
\includegraphics[width=\columnwidth]{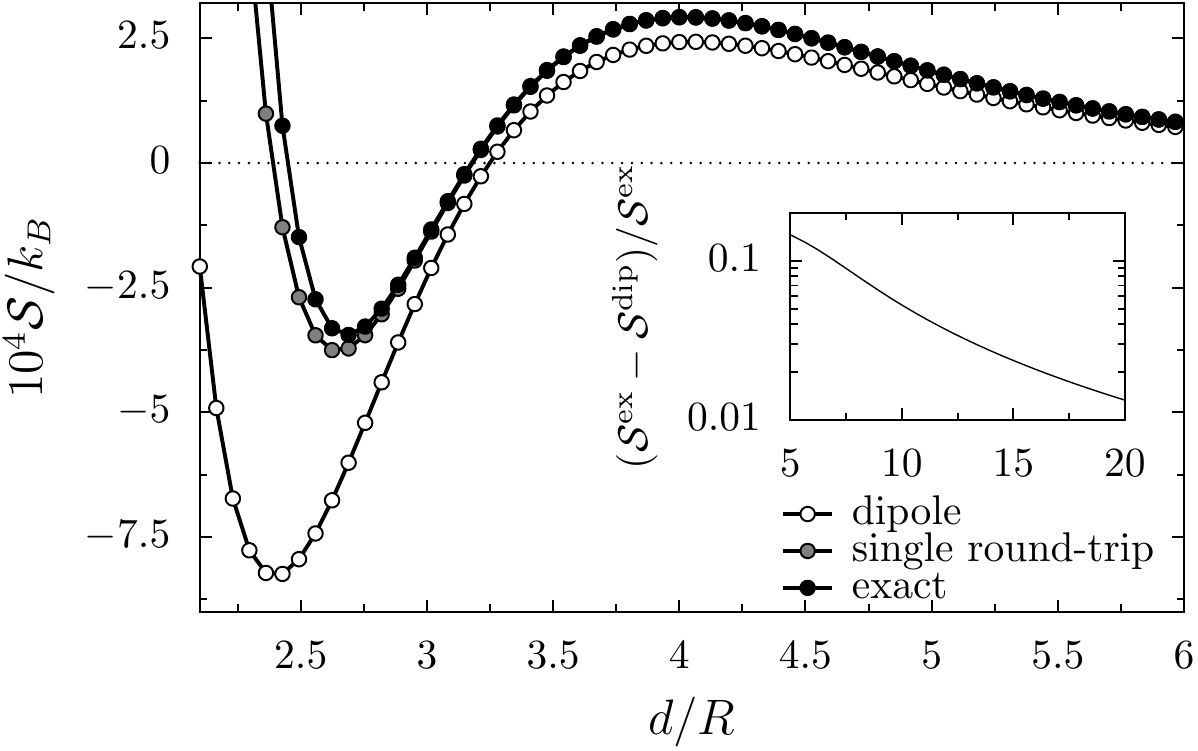}
\caption{The Casimir entropy $\mathcal{S}$ is presented as a function of the
  distance $d$ between two perfectly conducting spheres of radius $R$ at
  $2\pi k_BTR/\hbar c=1$ for different levels of approximation. Results
  obtained within the single round-trip approximation are displayed as white
  and grey points. While the first ones correspond to the dipole approximation, 
  the latter include higher multipoles up to $\ell_\text{max}=60$. The
  numerically exact result depicted by black points is obtained on the basis
  of (\ref{eq:free_energy}) with $\ell_\text{max} = 60$. The inset shows the 	
  relative difference of the dipole approximation $\mathcal{S}^{\rm dip}$ and
  the exact result $\mathcal{S}^{\rm ex}$.}
\label{fig:SRT_exact_PEC}
\end{figure}

Fig.~\ref{fig:SRT_exact_PEC} shows the Casimir entropy as a function of the
distance $d$ between two perfectly conducting spheres. The temperature is kept
fixed at $2\pi k_BTR/\hbar c=1$. The black points correspond to the full
calculation based on (\ref{eq:free_energy}) including multipoles up to
$\ell_\text{max}=60$ while the grey points correspond to the single
round-trip approximation (\ref{eq:single_scattering}). The white points are
obtained by additionally applying the dipole approximation.

The inset of Fig.~\ref{fig:SRT_exact_PEC} shows how the single round-trip
dipole approximation approaches the exact result for the Casimir entropy as
the distance between the two spheres is increased. At $d/R=20$, the distance
chosen in the previous sections, the relative error reaches about one percent.

At the temperature chosen here, the Casimir entropy is positive at large
distances between the two spheres. Decreasing the distance, the entropy will
eventually become negative and go through a minimum before rising again to
positive values \cite{Rodriguez2011}. Such a distance dependence was also found
for the plane-sphere geometry \cite{Canaguier2010,Canaguier2010a}.
Interestingly, even the single round-trip dipole approximation (white points)
is capable of qualitatively describing this distance dependence of the Casimir
entropy.

Including higher multipoles and multiple round-trips yields significant
positive contributions to the Casimir entropy in particular at relatively small
distances between the two spheres. However, already the single round-trip
approximation (grey points) provides a good description almost down to
distances where the minimum of the Casimir entropy is reached. Only at even
smaller distances multiple round-trip contributions become relevant (black
points).

\begin{figure}
\includegraphics[width=\columnwidth]{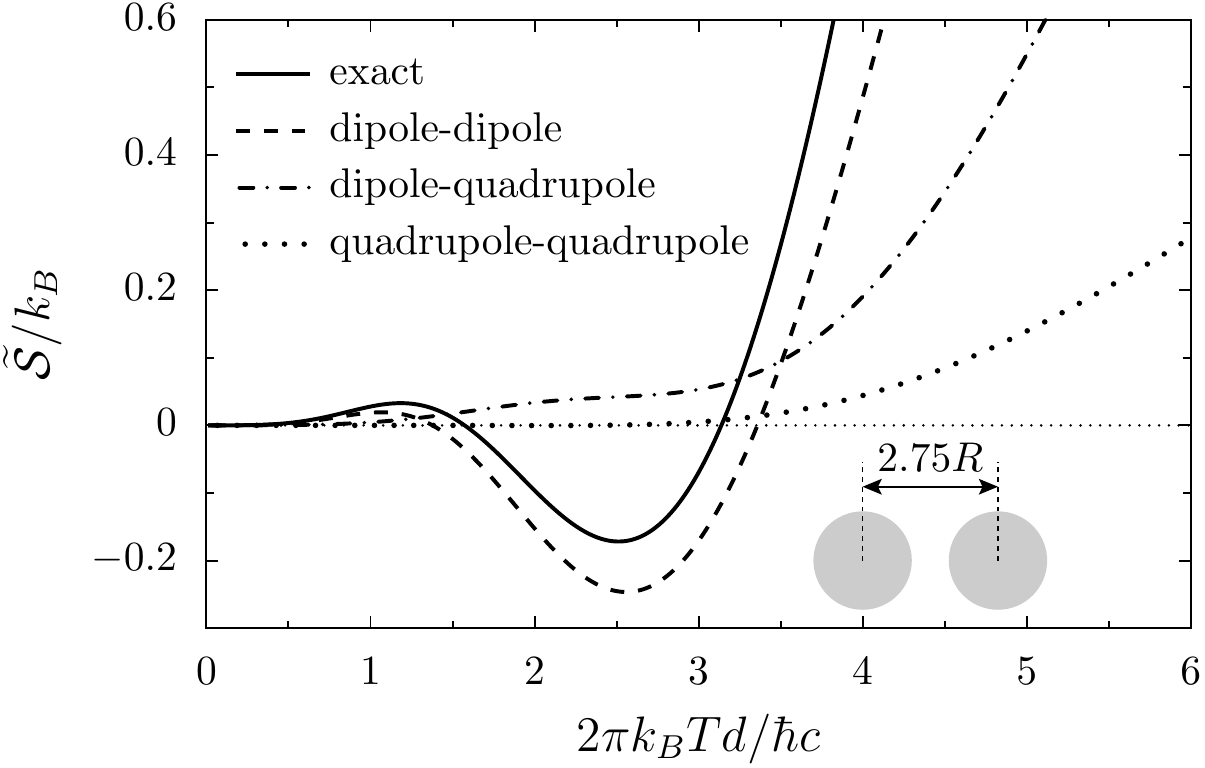}
\caption{The scaled Casimir entropy $\widetilde{\mathcal{S}} =
  (d/R)^6\mathcal{S}$ as a function of temperature is shown as solid line
  accounting for multipoles up to $\ell_\text{max}=15$. The contributions
  arising from dipole-dipole, dipole-quadrupole, and quadrupole-quadrupole
  scattering are depicted as dashed, dash-dotted, and dotted line, respectively.   The distance between the perfectly conducting spheres is $d=2.75R$.}
\label{fig:S_beyond_dipole_S_channels}
\end{figure}

Fig.~\ref{fig:SRT_exact_PEC} clearly demonstrates that for not too small
distances $d$ the main correction to the single round-trip dipole approximation
consists in the contribution of higher multipoles. We therefore analyze in
Fig.~\ref{fig:S_beyond_dipole_S_channels} how the dipole-quadrupole and
quadrupole-quadrupole channels contribute to the temperature dependence of the
Casimir entropy.  The distance between the two perfectly conducting spheres is
chosen as $d=2.75R$, i.e. close to the minimum of the Casimir entropy in
Fig.~\ref{fig:SRT_exact_PEC}.

The solid line depicts the Casimir entropy including multipoles up to
$\ell_\text{max}=15$ while the contributions due to dipole-dipole,
dipole-quadrupole, and quadrupole-quadrupole scattering are displayed by the
dashed, dash-dotted, and dotted lines, respectively, and comprise all
polarization channels. In agreement with Ref.~\cite{Ingold2015} we find that
the negative Casimir entropy is caused exclusively by dipole scattering. In
fact, as can be seen from Fig.~\ref{fig:S_beyond_dipole_S_channels} for
dipole-quadrupole and quadrupole-quadrupole scattering, higher multipoles lead
to positive Casimir entropy contributions. Still, an analysis of the
polarization channels involved in the multipole contributions depicted in
Fig.~\ref{fig:S_beyond_dipole_S_channels} reveals that they have the same signs
as the corresponding dipole-dipole contributions. However, for higher
multipoles the negative contribution from the polarization-mixing channel is
shifted to larger temperatures where the polarization-conserving channels
dominate, yielding an overall result with positive sign.

\begin{figure}
\includegraphics[width=\columnwidth]{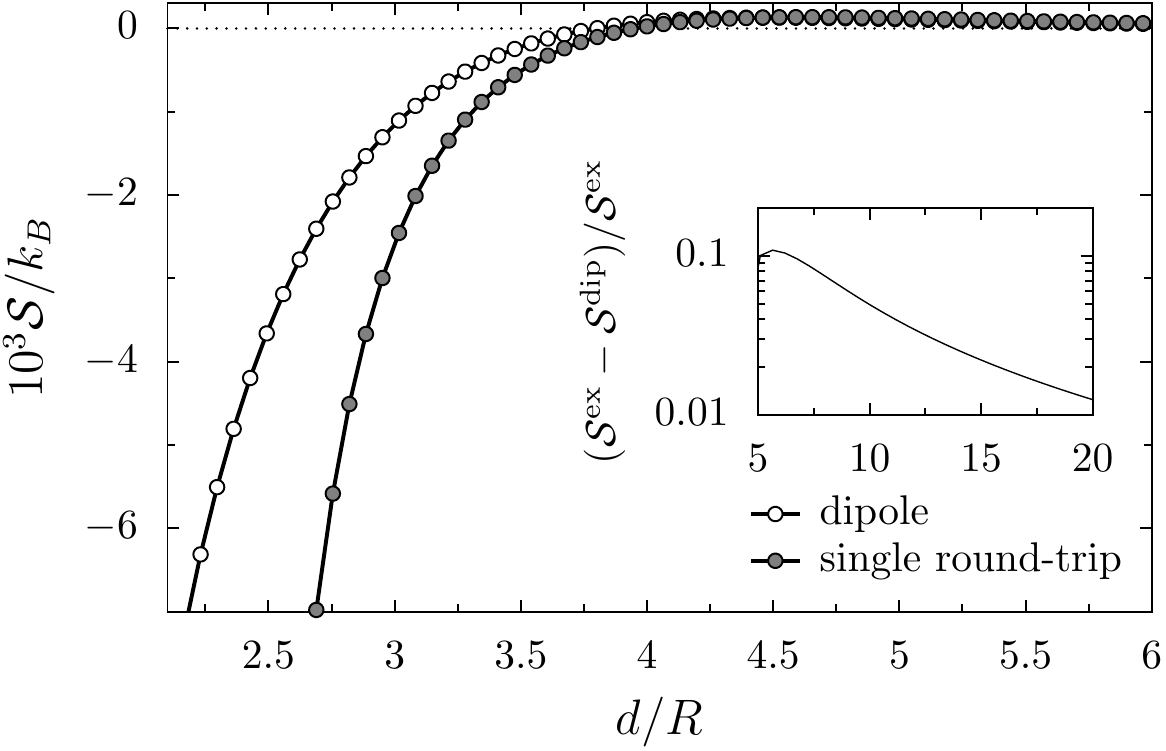}
\caption{The Casimir entropy $\mathcal{S}$ is presented as a function of the
  distance $d$ between Drude-type metal spheres of radius $R$ at
  $2\pi k_BTR/\hbar c=1$ within the single round-trip approximation. The
  material parameters correspond to those of the best conductor represented by
  a blue curve in Fig.~\ref{fig:PP_SP_SS_entropies}, i.e.
  $\omega_P R/2\pi c=20$ and $\omega_P/2\pi\gamma=4\times10^4$. The white
  points correspond to the dipole approximation while the grey points
  include higher multipoles up to $\ell_\text{max}=60$. Numerically exact
  results accounting for multiple round-trips cannot be distinguished from the
  grey points in this figure and are therefore not shown. The inset displays the
  relative difference of the dipole approximation $\mathcal{S}^{\rm dip}$ and
  the exact result $\mathcal{S}^{\rm ex}$.}
\label{fig:SRT_exact_Drude}
\end{figure}

We now turn to the dissipative case and consider Drude-type metal spheres. The
distance dependence of the Casimir entropy is depicted in
Fig.~\ref{fig:SRT_exact_Drude} for $2\pi k_BTR/\hbar c=1$. The material
parameters correspond to those of the best conductor considered in
Sec.~\ref{sec:dissgeo} and represented by the blue line in
Fig.~\ref{fig:PP_SP_SS_entropies}. Only results within the single round-trip
approximation are displayed in Fig.~\ref{fig:SRT_exact_Drude}, with white and
grey points representing, respectively, the dipole approximation and the
Casimir entropy with higher multipoles included. For the data presented here,
the numerically exact result accounting for multiple round-trips practically
coincides with the grey points.

In contrast to perfectly conducting spheres, the contributions arising from
higher multipoles are now negative and lead to a Casimir entropy increasing
monotonically with distance. While the geometrically induced negative Casimir
entropy is only due to dipole scattering, this is not the case for the negative
Casimir entropy caused by dissipation. Here, also higher multipole scattering
contributes with negative values. This important difference is to be expected
because dissipation can give rise to a negative Casimir entropy even in the
plane-plane geometry.

As the inset in Fig.~\ref{fig:SRT_exact_Drude} shows, the deviation of the
single round-trip dipole approximation from the numerically exact result is
comparable for, both, perfect conductor spheres and Drude-type metal spheres.
However, because of the missing high-temperature contribution of the TE modes,
the absolute value of the Casimir entropy in the dissipative case exceeds the
one for perfect conductors by one order of magnitude for the temperatures
considered here. As a consequence, multiple round-trip corrections are much less
important in Fig.~\ref{fig:SRT_exact_Drude} than they were in
Fig.~\ref{fig:SRT_exact_PEC}.  However, in both cases, the approximations
central to our analysis -- single round-trip and dipole scattering -- are well
justified for the large distances $d/R=20$ chosen in the previous sections.

\section{Conclusions}
\label{sec:conclusions}

Dissipation and scattering geometry represent possible sources of negative
Casimir entropies. While the first mechanism has been studied for quite some
time in the plane-plane geometry, the second mechanism was identified for
plane-sphere and sphere-sphere geometries with perfect conductors. For
Drude-type metals of sufficiently large dc conductivity, we have shown the
coexistence of both mechanisms by going beyond the Rayleigh limit.

In the large-distance limit, where only single round-trips are relevant, we
employed a scattering-channel analysis to disentangle dissipative and geometric
origins of the negative Casimir entropy. Concentrating on the sphere-sphere
geometry, we found that dissipation acts via the TE channel as in the
plane-plane geometry. In contrast, the geometric contribution can be traced
back to the polarization-mixing channel as in the perfectly conducting case.
Although being of different physical origin, both mechanisms are associated
with a vanishing free energy in the high-temperature limit.

Focussing on the geometrical origin of negative Casimir entropies, we have
found for perfectly conducting scatterers that while the dipole-dipole
channel can lead to a negative contribution to the Casimir entropy, higher
multipoles yield a positive contribution. With decreasing distance between the
scatterers, the higher multipoles become dominant and the total Casimir entropy
turns positive, thereby making connection with the plane-plane geometry for
perfect conductors. In contrast, for the dissipative mechanism leading to
negative Casimir entropies, the negative contribution is not restricted
to dipole scattering.

\begin{acknowledgments}
We are grateful to Astrid Lambrecht, Serge Reynaud, Antoine Canaguier-Durand,
Romain Gu{\'e}rout, and Kimball Milton for many insightful discussions. This
work has been supported by the DAAD and CAPES through the PROBRAL program.
In addition, PAMN acknowledges financial support by FAPERJ and CNPq (Brazilian
agencies).
\end{acknowledgments}

\appendix*
\section{Reflection and translation operators}
\label{appendix:Mie_scattering}

For the reflection at a sphere, the scattering operator $\mathcal{R}$ occurring
in (\ref{eq:roundtrip}) is naturally represented in the spherical multipole
basis characterized by $\ell$ and $m$. The corresponding matrix elements are
given by the Mie coefficients \cite{Bohren2007}. At imaginary frequencies and for
non-magnetic spheres, $\mu=1$, with refractive index $n=\sqrt{\varepsilon}$,
they can be expressed as
\begin{equation}
a_{\ell} =(-1)^{\ell+1}\frac{\pi}{2}
 \frac{n\psi_\ell(nkR)\psi_\ell'(kR)-\psi_\ell(kR)\psi_\ell'(nkR)}
 	  {n\psi_\ell(nkR)\xi_\ell'(kR)-\xi_\ell(kR)\psi_\ell'(nkR)}
\label{eq:mie_TM}
\end{equation}
for TM polarization
\begin{equation}
b_{\ell} = (-1)^{\ell+1}\frac{\pi}{2} 
 \frac{\psi_\ell(nkR)\psi_\ell'(kR)-n\psi_\ell(kR)\psi_\ell'(nkR)}
 	  {\psi_\ell(nkR)\xi_\ell'(kR)-n\xi_\ell(kR)\psi_\ell'(nkR)}
\label{eq:mie_TE}
\end{equation}
for TE polarization. The Riccati-Bessel functions are defined in terms of
modified spherical Bessel functions of the first kind and second kind
\cite{NistSpecial} as:
\begin{align}
 \psi_\ell(\rho) = \rho i_\ell(\rho),\ \ \xi_\ell(\rho) = \rho k_\ell(\rho).
\end{align}

The spherical vector wave basis refers to a coordinate system with origin in
the center of one of the spheres. A basis change from one center to another one
can be done by means of translation formulae
\cite{Stein1961,Cruzan1962,Bruning1971,Wittmann1988} which can also be
expressed in terms of imaginary wave vectors \cite{Emig2007}. The matrix
elements of the translation operator take a relatively simple form if the
translation occurs along the $z$ axis. Then, $m$ is conserved while $\ell$ and
the polarization P can change. For imaginary wave numbers $k$, the matrix
elements can be cast into the form
\begin{align}
\mathcal{T}^{\rm PP'}_{\ell_1,\ell_2;m}(kd) &=
\frac{(-1)^{m+1}(\pm\mathrm{i})^{\ell_1-\ell_2}}
      {\sqrt{\pi\ell_1(\ell_1+1)\ell_2(\ell_2+1)}}
\label{eq:TPPp}\\
&\hspace{-2em}\times\sum_{\ell'=|\ell_1-\ell_2|}^{\ell_1+\ell_2}
c_{\ell_1,\ell_2,\ell';m}^\text{PP'} Y^{\ell_1,\ell_2,\ell'}_{-m,m,0}
k_{\ell'}(kd)\,,\nonumber
\end{align}
where the initial and final polarizations P and P', respectively, may be equal
or different. The $\pm$ sign has to be interpreted as positive if the translation
occurs in the sense of the $z$ axis while the negative sign applies for
translations in the opposite direction.

If the polarization is conserved, $\text{P}=\text{P'}$, the
coefficient appearing in (\ref{eq:TPPp}) reads
\begin{equation}
c_{\ell_1,\ell_2,\ell';m}^\text{PP} = 
2\sqrt{2\ell'+1}[\ell_1(\ell_1+1)+\ell_2(\ell_2+1) -\ell'(\ell'+1)]
\label{eq:cpp}
\end{equation}
while for different polarizations we obtain
\begin{equation}
c_{\ell_1,\ell_2,\ell';m}^\text{PP'} = \pm4\sqrt{(2\ell'+1)}mkd
\ \ \ (\text{P}\neq\text{P'})\,.
\label{eq:cppp}
\end{equation}
In the latter case, the matrix elements of the translation operator vanish
in the limit of vanishing wave number $k$ as well as for $m=0$.

The Gaunt coefficients (see e.g. \cite{Rasch2003})
\begin{align}
\nonumber
Y^{\ell_1,\ell_2,\ell'}_{-m,m,0}
&=\sqrt{\frac{(2\ell_1+1)(2\ell_2+1)(2\ell'+1)}
                                         {4\pi}}\\
&\quad\times\left(
\begin{array}{ccc}
\ell_1 & \ell_2 & \ell' \\
0    &   0   &  0     
\end{array}
 \right)
\left(
\begin{array}{ccc}
\ell_1 & \ell_2 & \ell' \\
-m    &   m   &  0     
\end{array}
\right)
\label{eq:gaunt}
\end{align}
do not require the explicit evaluation of the two Wigner 3j symbols but can
be determined efficiently by means of recurrence relations \cite{Xu1997}.

In the dipole limit, $\ell_1=\ell_2=1$, the matrix elements of the translation
operator can be written as
\begin{align}
\label{eq:translation_PP_m0}
\mathcal{T}^{\rm PP}_{1,1;0}\mathcal{T}^{\rm PP}_{1,1;0} &= 
9e^{-2kd}\left(\frac{1}{(kd)^2}+\frac{1}{(kd)^3}\right)^2\,, \\
\label{eq:translation_PP_m1}
\mathcal{T}^{\rm PP}_{1,1;1}\mathcal{T}^{\rm PP}_{1,1;1} &= 
\frac{9}{4}e^{-2kd}\left(\frac{1}{kd}+\frac{1}{(kd)^2}+
\frac{1}{(kd)^3}\right)^2\,, \\
\label{eq:translation_PsP_m1}
\mathcal{T}^{\rm PP'}_{1,1;1}\mathcal{T}^{\rm P'P}_{1,1;1} &= 
-\frac{9}{4}e^{-2kd}\left(\frac{1}{kd}+\frac{1}{(kd)^2}\right)^2.		
\end{align} 
These expressions make the suppression of large wave numbers mentioned
in Sec.~\ref{sec:MieCoefficients} explicit.

\end{document}